\documentclass{elsart}

\usepackage{graphicx}
\usepackage{eucal}
\usepackage[dvips]{epsfig}
\usepackage{amssymb}
\usepackage{amsmath}
\usepackage{cite}

\begin{document}



\newcommand{\be}{\begin{eqnarray}}
\newcommand{\ee}{\end{eqnarray}}
\newcommand{\bse}{\begin{subequations}}
\newcommand{\ese}{\end{subequations}}

\newcommand{\bs}{\boldsymbol}
\newcommand{\mbb}{\mathbb}
\newcommand{\mcal}{\mathcal}
\newcommand{\mfr}{\mathfrak}
\newcommand{\mrm}{\mathrm}

\newcommand{\ovl}{\overline}
\newcommand{\p}{\partial}
\newcommand{\f}{\frac}
\newcommand{\diff}{\mrm{d}}
\newcommand{\lan}{\langle}
\newcommand{\ran}{\rangle}

\newcommand{\ga}{\alpha}
\newcommand{\gb}{\beta}
\newcommand{\gc}{\gamma}
\newcommand{\Gd}{\Delta}
\newcommand{\gd}{\delta}
\newcommand{\Gc}{\Gamma}
\newcommand{\gl}{\lambda}
\newcommand{\gk}{\kappa}
\newcommand{\go}{\omega}
\newcommand{\Go}{\Omega}
\newcommand{\Gs}{\Sigma}
\newcommand{\gs}{\sigma}
\newcommand{\veps}{\varepsilon}
\newcommand{\Gt}{\Theta}

\newcommand{\sn}{\mrm{sn}}
\newcommand{\cn}{\mrm{cn}}
\newcommand{\dn}{\mrm{dn}}
\newcommand{\am}{\mrm{am}}
\newcommand{\sech}{\mrm{sech}}
\newcommand{\sign}{\mrm{sign}}

\newcommand{\csp}{\;,\qquad\qquad}
\newcommand{\fa}{\forall\;}

\newcommand{\N}{\mbb{N}}
\newcommand{\R}{\mbb{R}}
\newcommand{\D}{\mcal{D}}
\newcommand{\Nn}{\mcal{N}}
\newcommand{\V}{\mcal{V}}

\newcommand{\im}{\mrm{image}\;}
\newcommand{\num}{\mrm{\#}}

\begin{frontmatter}

\title{One-dimensional nonrelativistic and relativistic Brownian motions: A microscopic collision model} 

\author{J\"orn Dunkel}
\ead{joern.dunkel@physik.uni-augsburg.de}
\ead[url]{www.physik.uni-augsburg.de/$\sim$dunkeljo}
\author{\and}
\author{Peter H\"anggi}
\ead{peter.hanggi@physik.uni-augsburg.de}
\ead[url]{www.physik.uni-augsburg.de/theo1/hanggi/}
\address{Institut f\"ur Physik, Universit\"at Augsburg,
 Theoretische Physik I,  Universit\"atsstra{\ss}e~1, D-86135 Augsburg, Germany}


\begin{abstract} 
We study a simple microscopic model for the one-dimensional stochastic motion of a (non)relativistic Brownian particle, embedded into a heat bath consisting of (non)relativistic particles. The stationary momentum distributions are identified self-consistently (for both Brownian and heat bath particles) by means of two coupled integral criteria. The latter follow directly from the kinematic conservation laws for the microscopic collision processes, provided one additionally assumes probabilistic independence of the initial momenta. It is shown that, in the nonrelativistic case, the integral criteria do correctly identify the Maxwellian momentum distributions as stationary (invariant) solutions. Subsequently, we apply the same criteria to the relativistic case. Surprisingly, we find here that the stationary momentum distributions differ slightly from the standard J\"uttner distribution by an additional prefactor proportional to the inverse relativistic kinetic energy. 
\end{abstract}
\end{frontmatter}

\section{Introduction}
\label{introduction}

The implementation of the Brownian motion concept 
\cite{Ei05c,EiSm,UhOr30,Ch43,WaUh45,KaSh91} into special relativity
\cite{Ei05a,Ei05b} represents a longstanding
issue in mathematical  and statistical physics (classical references
are \cite{Sc61,Ha65,Du65}; more recent contributions include
\cite{GuRu78,Bo79a,Roy80,BY81,MoVi95,Po97,Ra05,Ra05a,OrHo05,FrLJ06,Fi06}; for
a kinetic theory approach, see \cite{DG80,CeKr02,BaMo00}). In two
recent papers \cite{DuHa05a,DuHa05b} we have discussed in detail how one can construct
Langevin equations for relativistic Brownian motions (see 
Debbasch et al. \cite{DeMaRi97,De04} and Zygadlo \cite{Zy05} for similar approaches, and also Dunkel and H\"anggi~\cite{DuHa06b}).  Thereby, 
it was  demonstrated that, in general, the relativistic Langevin
equation {\it per se} cannot uniquely determine the corresponding
Fokker-Planck equation (FPE). This dilemma is caused by the fact
that the relativistic Langevin equations, e.g., if written in
laboratory coordinates, may exhibit a multiplicative coupling between a
function of the momentum coordinate and a Gaussian white noise process
(laboratory frame $\equiv$ rest frame of the heat bath). Thus,
depending on the choice of the discretization rule, one
obtains different forms of relativistic FPEs characterized by different
stationary solutions.      
\par
In Refs. \cite{DuHa05a,DuHa05b} we have analyzed the three most popular discretization
rules for Langevin equations with multiplicative noise, namely, 
Ito's pre-point discretization rule \cite{Ito44,Ito51}, the
Fisk-Stratonovich mid-point rule 
\cite{St64,St66,Fisk63,Fisk65}, and the H\"anggi-Klimontovich
(HK) post-point rule \cite{Han78,Han80,HaTo82,Kl94}. As a main result it was found that
only the HK interpretation of the Langevin equation yields a FPE,
whose stationary solution coincides with the one-dimensional
relativistic  J\"uttner-Maxwell distribution, as known from J\"uttner's early
work on the relativistic gas \cite{Ju11,Sy57} and also from the
relativistic kinetic theory \cite{DG80,CeKr02}.  Thus, in absence of
other qualifying criteria, one may conlude that the post-point
discretization rule is favorable. However, it naturally arises the
question, if one can gain additional insights by studying the
microscopic collision processes, which cause the stochastic motion of a Brownian particle.
\par
The present paper intends to partially address this question for the one-dimensional (1D) case; i.e.,  instead of focusing on the \lq macroscopic\rq\space Langevin formulation, we will study here Brownian motions by means of a simple microscopic 
1D model. Since, on the level of the  Brownian motion
approach, the exact nature of the microscopic interaction forces is
usually negligible we shall restrict ourselves to purely elastic
collisions between a Brownian particle and the constituents of a
surrounding heat bath. As will be demonstrated below, this simple model suffices to identify stationary (invariant)
momentum distributions for both non-relativistic and relativistic 1D
collision processes~\footnote{More precisely, we shall additionally
require the probabilistic independence of the initial momenta during each
single collision -- but this is a rather reasonable, weak
restriction.}.
\par 
The paper is organized as follows: Section \ref{s:non-relativistic} is
dedicated to the non-relativistic case, serving as the test
example for our approach. After briefly summarizing the basic model assumptions, we
will derive two coupled integral criteria (Sec. \ref{s:integral}),
which can be used to identify the invariant momentum distributions for
the heat bath particles and the Brownian particles, respectively. It is shown that these integral criteria do
indeed yield the correct stationary momentum distribution of the
non-relativistic Brownian motion, namely, the well-known Maxwellian
momentum distribution (Sec. \ref{s:Maxwell}). Afterwards, the same method is applied to the relativistic case (Sec. \ref{s:relativistic}). Remarkably, we
find that, under exactly same preconditions, the invariant relativistic distributions do {\em not} exactly correspond to the standard J\"uttner distributions, but rather to modified  J\"uttner functions including an additional prefactor $1/E$, where $E$ is the relativistic kinetic energy of the particle under consideration.  The paper concludes with a summary and a discussion of the results (Sec. \ref{summary}). 

\section{Non-Relativistic Brownian motions}
\label{s:non-relativistic}

In this part, we review non-relativistic 1D-Brownian
motions. Later on, we will pursue an analogous approach to identify
the stationary momentum distribution of a relativistic Brownian
particle, embedded into a relativistic heat bath.

\subsection{Basic model assumptions}
\label{basic-nr}

In the three-dimensional (3D) case a simple idealized model for
Brownian motions can be imagined as follows: Consider a (infinitely
heavy) box of volume $\V$, possessing diathermal \cite{MaKrSz05} walls
and being at rest in the inertial laboratory frame $\Gs_0$. Let this
box contain a homogeneous, quasi-ideal (weakly interacting)  gas,
consisting of approximately point-like particles with identical masses
$m$. Further, assume that the gas (or liquid) particles -- referred to
as \lq heat bath\rq, hereafter -- surround a  Brownian particle of
mass $M\gg m$. Then, 
due to frequent elastic collisions with heat bath particles, the Brownian particle performs 3D random motions. Given the distribution of the heat bath, the stochastic dynamics of the Brownian particle is determined by the collision kinematics (cross-sections) governing the interaction with the heat bath particles. 
\par  
If, as in the present paper, one wishes to study the 1D case, slight
modifications of the above model are necessary. The reason is that,
typically, two particles cannot simply exchange positions if their
motions are confined to one dimension. To circumvent this problem, we
shall therefore imagine the heat bath particles as having \emph{fixed
positions} on a 1D lattice but \emph{non-vanishing
momenta}. Correspondingly, in this 1D (lattice) model the Brownian
particle may jump from one lattice point to the next during one time
step. Additionally, we will impose that at each lattice point there
does indeed occur an elastic interaction in accordance with the laws
of momentum and energy conservation. Mathematically, the latter
assumption corresponds to considering distributions conditional on the
event \lq\lq a collision has occurred\rq\rq. 
\par
Based on this idealized 1D (lattice) model, it is our primary
objective to determine self-consistently the invariant (i.e., stationary) momentum distributions for both heat bath and
Brownian particles. To this end, we shall next summarize the simple kinematic equations governing the collisions in this model. By interpreting the momentum coordinates as coupled random variables, we then derive in  Sec.~\ref{s:integral} two general integral criteria which have to be satisfied by the stationary distributions. As shown in Sec.~\ref{s:Maxwell}, in the non-relativistic case the stationary solutions are given by the well-known Maxwell distributions. The integral criteria apply to both non-relativistic and relativistic collisions; hence, we can use them later to also identify the stationary distributions of the corresponding relativistic model (cf. Sec.~\ref{s:relativistic}).

\subsection{Kinematics of single collision events}
\label{s:non-relativistic_kinematics}

The momentum and energy balance per (elastic) collision reads
\bse
\be
E+\epsilon&=&\hat{E}+\hat{\epsilon},\\
P+p&=&\hat{P}+\hat{p}.
\ee
\ese
Here and below, capital letters refer to the Brownian particle and
small letters to heat bath particles; quantities without
(with) hat-symbols refer to the state before (after) the
collision. In the non-relativistic case, we have, e.g., before the collision 
\bse\label{e:non-relativistic_definitions}
\be
P=MV,\qquad p=mv,\\
E=\f{P^2}{2M},\qquad \epsilon=\f{p^2}{2m},
\ee
\ese
where $v$ and $V$ denote the velocities with respect to the laboratory frame $\Sigma_0$. Taking into account both conservation of momentum and (kinetic) energy, one finds for a single collision the elementary results 
\bse\label{e:p-tilde-nonrel}
\be
\hat{P}(p,P)&=&\label{e:p-tilde-nonrel-a}
\left(\f{2M}{M+m}\right) \,p+\left(\f{M-m}{M+m}\right)\,P,\\
\hat{p}(P,p)&=&\label{e:p-tilde-nonrel-b}
\left(\f{2m}{M+m}\right)\,P+\left(\f{m-M}{M+m}\right) \,p.
\ee  
\ese
We again stress that Eqs. \eqref{e:p-tilde-nonrel} do implicitly assume that a  collision has indeed occurred (otherwise, the momenta would remain unchanged); i.e., in mathematical terms, any results obtained by employing Eqs. \eqref{e:p-tilde-nonrel} are valid conditional on the information that a collision event has taken place. 
\par 
Now let us suppose we know the joint two-particle PDF $\psi_2(p,P)$ for the particle momenta \emph{before} the collision. Then, the kinematic laws \eqref{e:p-tilde-nonrel} determine  uniquely the marginal momentum PDFs $\hat{\Phi}(\hat{P})$ and $\hat{\phi}(\hat{p})$ after the collision, formally defined by
\bse
\be
\hat{\Phi}(\hat{P})&=&\int \diff \hat p\; \hat{\psi}_2(\hat p,\hat P),\\
\hat{\phi}(\hat{p})&=&\int \diff \hat P\; \hat{\psi}_2(\hat p,\hat P),
\ee 
\ese
where $\hat{\psi}_2(\hat p,\hat P)$ is the joint momentum PDF \emph{after} the collision 
(here and below integrals with unspecified boundaries range from
$-\infty$ to $+\infty$.)  Our main objective will be to identify
\emph{stationary (invariant)} momentum distributions,  satisfying  \emph{by definition}
\be\label{e:stationarity_criterion_0}
\hat{\psi}_2(\hat p,\hat P)\equiv {\psi}_2(\hat p,\hat P).
\ee
The latter condition just means that a stationary PDF must remain
invariant in  microscopic collisions. In particular look, we shall for
particular stationary solutions which can be written in the product form
\be\label{e:product_ansatz}
{\psi}_2(p,P)=\phi(p)\,\Phi(P).
\ee
Mathematically, this corresponds to the assumption that, \emph{in the stationary state, the momenta $P$ and $p$ can be considered as independently distributed random variables}. 
For stationary PDFs of the form \eqref{e:product_ansatz} the stationarity criterion \eqref{e:stationarity_criterion} reduces to
\be\label{e:stationarity_criterion}
\hat\Phi(\hat P)=\Phi(\hat P),\qquad
\hat\phi(\hat p)=\phi(\hat p),
\ee 
where $\hat{\psi}_2(\hat p,\hat P)=\hat\phi(\hat p)\,\hat\Phi(\hat P)$ is the joint distribution after the collision.  From a physical point of view, the independence assumption for $(p,P)$ or, alternatively, of  $( \hat p, \hat P)$ is guided by the experience that the well-known equilibrium momentum distributions of quasi-ideal  non-relativistic and relativistic $N$ particle gases (i.e. the Maxwell and  J\"uttner distributions) can be written as products of one-particle momentum distributions. 
\par
In order to be able to determine the stationary PDFs for a given collision kinematics, we next derive general integral criteria. It comes as no surprise that, in the non-relativistic case, the stationary solutions will be given by a pair of Maxwellians. 

\subsection{Integral criteria for stationary momentum distributions} 
\label{s:integral}

Consider two \emph{independently} distributed random variables $Y$ and
$Z$, and a derived random variable $X=X(Y,Z)$. The corresponding PDFs
are denoted by $\Phi_X(X)$, $\Phi_Y(Y)$ and $\Phi_Z(Z)$. The average
of some test function $g(X)$ with respect to $\Phi_X$ can then be written as 
\be
\int \diff X\; g(X)\; \Phi_X(X)
=
\int \diff Y\int \diff Z\; g(X(Y,Z))\; \Phi_Y(Y)\,\Phi_Z(Z).
\ee
Assuming that the (partially) inverse transformation $Z=Z(Y,X)$ is well-defined  (i.e., strictly monotonous for each fixed $Y$), and that Fubini's theorem~\cite{Gr02} is applicable, we can rewrite the last equation in the form
\be
\int \diff X\; g(X)\; \Phi_X(X)
&=&\notag
\int \diff Y\; \int \diff X\,\left|\f{\p Z}{\p X}\right|g(X)\;
\Phi_Y(Y)\,\Phi_Z(Z(Y,X))\\
&=&
\int \diff X\;g(X)\int \diff Y\,\left|\f{\p Z}{\p X}\right|\;
\Phi_Y(Y)\,\Phi_Z(Z(Y,X)).
\ee
Since the latter equation holds for any test function $g$, one obtains the well-known transformation law
\be\label{e:stationarity-1}
\Phi_X(X)=
\int \diff Y\, \left|\f{\p Z}{\p X}\right|\; \Phi_Y(Y)\;\Phi_Z(Z(Y,X)).
\ee
For completeness, we note that Eq.~\eqref{e:stationarity-1} can also be obtained by starting from  
\be
\Phi_X(x)=\int \diff Y \int \diff Z\, \gd(x-X(Y,Z))\;\Phi_Y(Y)\;\Phi_Z(Z)
\ee
and performing the $Z$-integration (with $\gd$ denoting the Dirac
delta-function). We next consider an explicit example, which will be
investigated in detail in the remainder.

\paragraph*{Example: $\hat{P}=\hat{P}(p,P)$ and $\hat{p}=\hat{p}(P,p)$.} 
The idea is that we express the final momenta in terms of the initial
momenta, cf. Eq.~\eqref{e:p-tilde-nonrel}. That is, setting $X=\hat{P}$, $Y=p$, $Z=P$ and, correspondingly,  $\Phi_X\equiv \hat\Phi$, $\Phi_Y\equiv\phi$, $\Phi_Z\equiv\Phi$, we can write Eq.~\eqref{e:stationarity-1} as
\bse\label{e:nonstationary}
\be\label{e:nonstationary-a}
\hat\Phi(\hat{P})=\int \diff p\,\biggl|\f{\p P}{\p \hat{P}}\biggr|\,\phi(p)\;
\Phi(P(p,\hat{P})).
\ee
For a given pair of initial distributions $(\Phi,\phi)$, this equation can be used to calculate the PDF of the Brownian particle after the collision. Analogously, by setting $X=\hat{p}$, $Y=P$, $Z=p$ and $\Phi_X\equiv \hat\phi$, $\Phi_Y\equiv\Phi$,$\Phi_Z\equiv\phi$, we find for the PDF of the
heat bath particles
\be\label{e:nonstationary-b}
\hat\phi(\hat{p})=\int \diff P\;\biggl|\f{\p p}{\p \hat{p}}\biggr|\,\Phi(P)\;
\phi(p(P,\hat{p})).
\ee
\ese
In particular, for \emph{stationary} distributions satisfying Eqs.~\eqref{e:stationarity_criterion} we have $\hat\phi\equiv \phi$ and $\hat\Phi\equiv \Phi$  and, hence, obtain from Eqs.~\eqref{e:nonstationary} the integral criteria 
\bse\label{e:stationarity-2}
\be
\Phi(\hat{P})&=&\label{e:stationarity-2a}
\int \diff p\,\biggl|\f{\p P}{\p \hat{P}}\biggr|\,\phi(p)\;
\Phi(P(p,\hat{P})),\\
\phi(\hat{p})&=&\label{e:stationarity-2b}
\int \diff P\;\biggl|\f{\p p}{\p \hat{p}}\biggr|\,\Phi(P)\;
\phi(p(P,\hat{p})).
\ee
\ese
Given a certain microscopic kinematic law, any pair $(\Phi,\phi)$
satisfying Eqs.~\eqref{e:stationarity-2} provides a self-consistent
stationary distribution. For completeness we mention that
mathematically equivalent criteria are obtained by exchanging the
positions of $p$ and $P$ as functional arguments, i.e., by considering
$\hat{P}=\hat{P}(P,p)$ and/or $\hat{p}=\hat{p}(p,P)$, respectively. 
\par
Before discussing solutions of Eqs.~(\ref{e:stationarity-2}) for the non-relativistic Brownian motion, it is worthwhile to stress the following fact: Since the derivation
of  Eqs.~(\ref{e:stationarity-2}) is based on rather
general assumptions, these integral equations can be applied to find the
stationary PDF $\Phi$ not only in the non-relativistic but also in the
relativistic case (as will be done in Sec. \ref{s:relativistic}). The
additional mathematical assumption, underlying the derivation of
Eqs.~(\ref{e:stationarity-2}), is that the initial momenta $P$ and $p$
can be viewed as  independently distributed random variables; i.e,
loosely speaking, this postulate is the only point leaving some
freedom for potential modifications, all other parts are dictated by
the physical conservation laws.  As stated before, from a physical
point of view, the independence assumption for $(p,P)$ or,
alternatively, of $(\hat p, \hat P)$ is guided by the experience that
the well-known equilibrium momentum distributions of quasi-ideal
non-relativistic and relativistic $N$ particle gases (i.e. the Maxwell
and  J\"uttner distributions) can usually be written as products of one-particle momentum distributions. 

\subsection{Stationarity of the Maxwell distribution} 
\label{s:Maxwell}

In the last part of this section, we briefly outline that the integral criteria Eqs.~\eqref{e:stationarity-2} are satisfied by the normalized Maxwell distributions 
\bse\label{e:Maxwell}
\be
\Phi(P)&=&\label{e:Maxwell_BP}
\left(\f{1}{2\pi Mk_B T}\right)^{1/2}
\exp\biggl(-\f{P^2}{2M k_B T}\biggr),\\
\phi(p)&=&\label{e:Maxwell_HBP}
\left(\f{1}{2\pi mk_BT}\right)^{1/2}
\exp\biggl(-\f{p^2}{2m k_B T}\biggr),
\ee
\ese
where $k_B$ denotes the Boltzmann constant, and $T$ is the temperature parameter. 
\par
In order to apply Eqs.~\eqref{e:stationarity-2}, we require functions  $P(p,\hat{P})$ and $p(P,\hat{p})$. From Eqs.~\eqref{e:p-tilde-nonrel}, we find
\bse\label{e:p-inv}
\be
P(p,\hat{P})&=&
\left(\f{M+m}{M-m}\right)\,\hat{P}-\left(\f{2M}{M-m}\right) \,p\\
p(P,\hat{p})&=&
\left(\f{M+m}{m-M}\right)\,\hat{p}-\left(\f{2m}{m-M}\right) \,P,
\ee
\ese
and, thus, Eqs. \eqref{e:stationarity-2} take the explicit form 
\bse\label{e:stationarity-3}
\be
\Phi(\hat{P})&=&
\left|\f{M+m}{M-m}\right|
\int \diff p\;\phi(p)\,\Phi\bigl(P(p,\hat{P})\bigr)\\
\phi(\hat{p})&=&
\left|\f{M+m}{m-M}\right|
\int \diff P\;\Phi(P)\,\phi\bigl(p(P,\hat{p})\bigr).
\ee
\ese
As one can now easily verify by insertion, the Maxwell distributions \eqref{e:Maxwell} do indeed satisfy Eqs. \eqref{e:stationarity-3}. Consequently, Eqs.~\eqref{e:Maxwell} provide a pair of self-consistent stationary solutions for the non-relativistic collision kinematics, conditional on the information that a collision has  occurred.

\section{Relativistic Brownian motions}
\label{s:relativistic}

Since the integral equations \eqref{e:nonstationary} and
\eqref{e:stationarity-2} applied to the relativistic case as well, we
can, in principle, proceed exactly analogous to the non-relativistic
case. However, some purely technical difficulties arise due to the facts that (i) the relativistic collision kinematics is more complex than the non-relativistic one, and that (ii) the potential candidates for stationary distributions  do not allow for the solving Eqs. \eqref{e:nonstationary} and \eqref{e:stationarity-2} analytically.  
Hence, after having specified all required transformation formulae (Sec. \ref{s:relativistic_kinematics}), we
will evaluate the PDFs on the left-hand-sides of the non-stationary
integral equations   \eqref{e:nonstationary} numerically \footnote{Since the integrals
\eqref{e:nonstationary} are one-dimensional they can be solved
numerically with e.g. the software package {\sc
Mathematica}~\cite{Ma03}.}, probing
different types of candidate distributions
(Sec. \ref{s:relativistic_criterion}).

\subsection{Kinematics of a single collision event}
\label{s:relativistic_kinematics}

In the relativistic case, the momentum and energy balance per (elastic) collision can again be written in the form 
\bse\label{e:rel_kinematics} 
\be
E+\epsilon&=&\hat{E}+\hat{\epsilon},\\
P+p&=&\hat{P}+\hat{p}.
\ee
\ese
Compared with the non-relativistic case, the only difference is that we now use the relativistic expressions for momentum and kinetic energy, respectively. Specializing to the laboratory frame $\Gs_0$ (=rest frame of the heat bath), and using units such that the speed of light $c=1$, we have  [compare Eqs. \eqref{e:non-relativistic_definitions}]
\bse\label{e:relativistic_definitions}
\begin{align}
P&=MV\;\gc(V), &p&=mv\;\gc(v),\\
E&=(M^2+P^2)^{1/2}, &\epsilon&=(m^2+p^2)^{1/2},
\end{align}
where $V$ and $v$ denote the particles' velocities, and 
\be
\gc(v)=\left(1-v^2\right)^{-1/2}.
\ee
\ese
Suppose we are given the information that a single collision has occurred. Then, 
solving Eqs. \eqref{e:rel_kinematics} for $\hat{P}$, we find the explicit 
representations [compare Eqs.~\eqref{e:p-tilde-nonrel}] 
\bse\label{e:p-tilde-rel}
\be
\hat{P}(p,P)&=&\f{2v_0E-(1+v_0^2)P}{1-v_0^2},\label{e:p-tilde-rel-a}\\
\hat{p}(P,p)&=&\f{2v_0\epsilon-(1+v_0^2)p}{1-v_0^2},\label{e:p-tilde-rel-b}
\ee
where the velocity
\be
v_0=\f{p+P}{\epsilon+E}=\f{\hat{p}+\hat{P}}{\hat{\epsilon}+\hat{E}}=\hat{v}_0
\ee
\ese
corresponds to the Lorentz boost from $\Gs_0$ to the center-of-mass frame (see App.~\ref{a:relativistic_collisions} for details of the calculation). As one may easily check, in the non-relativistic limit case Eqs.~\eqref{e:p-tilde-rel} reduce to  Eqs.~\eqref{e:p-tilde-nonrel}.
\par
In order to be able to apply the integral criteria
\eqref{e:nonstationary}, we also need to determine the (partially)
inverse transformations $P(p,\hat{P})$ and $p(P,\hat{p})$,
respectively. In the non-relativistic case, this task was rather
simple, see Eqs.~\eqref{e:p-inv}. In the relativistic case, however,
a bit of extra care is required. To illustrate this, let us first 
consider the momentum equation \eqref{e:p-tilde-rel-a} for Brownian
particle. For any fixed value $p$ with $|p|<\infty$, one finds 
\bse\label{e:limits-1}
\be
\hat{P}_+(p)&:=&\lim_{P\to +\infty} \hat P(p,P)
=\f{(m^2+M^2)\;p+(M^2-m^2)\;\epsilon}{2m^2},\\
\hat{P}_-(p)&:=& \lim_{P\to -\infty} \hat P(p,  P)
=\f{(m^2+M^2)\;p-(M^2-m^2)\;\epsilon}{2m^2}.
\ee
\ese
Hence, at finite $p$, the inverse transformation $P(p,\hat{P})$ has a
finite support, corresponding to the interval (without loss of
generality, we will assume here and below that $M>m$)  
$$
I(p)=[\hat{P}_-(p),\hat{P}_+(p)].
$$
Hence, we find the following explicit form of the inverse transformation:  
\bse\label{e:p-tilde-rel-inv-1}
\be
P(p, \hat P) =\f{Q+R}{S},\qquad \hat{P}\in I(p),\quad p\in(-\infty,\infty),
\ee
with abbreviations
\be
Q&=&[m^2-M^2+2\,(p-\hat{P})\,p]\;[M^2 p-m^2\hat{P} +M^2\,(p-\hat{P})]\,,\\
R&=&2\,(p-\hat{P})\;(m^2-M^2)\;\epsilon\;\hat{E}\,,\\
S&=&(m^2-M^2)^2-4(p-\hat{P})(M^2p-m^2\hat P)\,.
\ee
\ese
Note that the limits in Eqs.~\eqref{e:limits-1} correspond to the
curves $S=0$. For later use, we also give the formal inversion of
Eqs.~\eqref{e:limits-1}:  
\bse\label{e:limits-1-inverted}
\be
p_+(\hat{P})&:=&
\f{(m^2+M^2)\;\hat{P}+(M^2-m^2)\;\hat{E}}{2M^2},\\
p_-(\hat{P})&:=&
\f{(m^2+M^2)\;\hat{P}-(M^2-m^2)\;\hat{E}}{2M^2},
\ee
\ese
which allow us to rewrite Eq.~\eqref{e:p-tilde-rel-inv-1} equivalently as
\be\label{e:p-tilde-rel-inv-1-alt}
P(p, \hat P) =\f{Q+R}{S},\qquad \hat{P}\in(-\infty,\infty),\quad
p\in [p_-(\hat{P}),p_+(\hat{P})].
\ee
Equation \eqref{e:p-tilde-rel-inv-1-alt} is in such a form that it can
directly be inserted into the integral equation
\eqref{e:nonstationary-a}.  
\par
Finally, going through an analogous analysis for the heat bath particle yields
\bse\label{e:p-tilde-rel-inv-2}
\be
p(P, \hat p)=\f{q+r}{s}, 
\qquad \hat{p}\in [\hat{p}_-(P),\hat{p}_+(P)],\quad P\in(-\infty,\infty),
\ee
where
\be
q&=&[M^2-m^2+2\,(P-\hat{p})\,P]\;[m^2 P-M^2\hat{p} +m^2\,(P-\hat{p})]\,,\\
r&=&2\,(P-\hat{p})\;(M^2-m^2)\;\hat{\epsilon}\;E\,,\\
s&=&(m^2-M^2)^2-4(P-\hat{p})(m^2P-M^2\hat p)\,,
\ee
\ese
and 
\bse\label{e:limits-2}
\be
\hat{p}_+(P)&:=&\lim_{p\to +\infty} \hat p(P, p)
=\f{(m^2+M^2)\;P +(M^2-m^2)\;E}{2M^2},\\
\hat{p}_-(P)&:=&\lim_{p\to -\infty} \hat p(P, p)
=\f{(m^2+M^2)\;P -(M^2-m^2)\;E}{2M^2}.
\ee
\ese
Note that the limits in Eq.~\eqref{e:limits-2} correspond to the
curves $s=0$, and their formal inversions can be defined by 
\bse\label{e:limits-2-inverted}
\be
{P}_+(\hat p)&:=&
\f{(m^2+M^2)\;\hat{p}+(M^2-m^2)\;\hat \epsilon}{2m^2},\\
{P}_-(\hat p)&:=& 
\f{(m^2+M^2)\;\hat p-(M^2-m^2)\;\hat \epsilon}{2m^2},
\ee
\ese
which allows us to rewrite Eq.~\eqref{e:p-tilde-rel-inv-1} equivalently as
\be\label{e:p-tilde-rel-inv-2-alt}
p(P, \hat p) =\f{q+r}{s},\qquad \hat{p}\in(-\infty,\infty),\quad
P\in [P_-(\hat{p}),P_+(\hat{p})].
\ee
Equation \eqref{e:p-tilde-rel-inv-2-alt} is given in such a form that it can
directly be inserted into the integral equation
\eqref{e:nonstationary-b}.  

\subsection{Testing the integral criterion}
\label{s:relativistic_criterion}

Analogous to the non-relativistic case, we aim to exploit the integral criteria \eqref{e:nonstationary} in order to determine the invariant momentum distributions. To this end, we proceed as follows: For functions $P(\hat{p},\hat P)$ and $p(\hat{P},\hat p)$, given in Eqs.~\eqref{e:p-tilde-rel-inv-1-alt} and \eqref{e:p-tilde-rel-inv-2-alt}, we evaluate numerically the integrals \eqref{e:nonstationary}: 
\bse\label{e:criteria_relativistic_used}
\be
\hat\Phi(\hat{P})=\int_{p_-(\hat{P})}^{p_+(\hat{P})}\diff p\,\biggl|\f{\p P}{\p \hat{P}}\biggr|\,\phi(p)\;
\Phi(P(p,\hat{P})),\\
\hat\phi(\hat{p})=\int_{{P}_-(\hat p)}^{{P}_+(\hat p)} 
\diff P\;\biggl|\f{\p p}{\p \hat{p}}\biggr|\,\Phi(P)\;
\phi(p(P,\hat{p})).
\ee
\ese
for a number specified grid-points $\hat P$ and $\hat{p}$,
respectively. The boundaries of the integration are chosen in
accordance with the support intervals of the functions $P(\hat{p},\hat
P)$ and $p(\hat{P},\hat p)$; cf. Eqs.~\eqref{e:p-tilde-rel-inv-1-alt} and
Eqs.~\eqref{e:p-tilde-rel-inv-2-alt}, respectively. We use different pairs of initial PDFs $(\Phi,\phi)$, and test if 
\be\label{e:stationary-sim}
\hat{\Phi}(\hat{P})\equiv \Phi(\hat P),\qquad
\hat\phi(\hat{p})\equiv \phi(\hat p)
\ee
hold simultaneously. If the answer is positive, we conclude that the candidate functions $(\Phi,\phi)$ fulfill the stationarity criteria \eqref{e:stationarity-2} and, thus, are invariant solutions for the relativistic collision process.  
\par 
We next discuss our choice of the initial PDFs. Guided by the results of \cite{DuHa05a,DuHa05b}, we consider   
\bse\label{e:Rel_Maxwell}
\be
\Phi(P)&=&\label{e:Rel_Maxwell_BP}
\f{\mcal{N}_\eta(M)}{E^\eta}
\exp\biggl(-\f{E}{k_B T}\biggr),\\
\phi(p)&=&\label{e:Rel_Maxwell_HBP}
\f{\mcal{N}_\eta(m)}{\epsilon^\eta}
\exp\biggl(-\f{\epsilon}{k_B T}\biggr),
\ee
\ese
where $\eta\in[0,1]$ is a free parameter, and  $\epsilon$ and $E$ denote the relativistic kinetic energies, respectively. By choosing the symmetric candidate distributions \eqref{e:Rel_Maxwell}, we automatically specialize to the rest frame of the heat bath. 
For a given set of parameters $(m,M,T,\eta)$ the normalization constants $\mcal{N}_\eta(m)$ and $\mcal{N}_\eta(M)$ are determined by the conditions
\be
1=\int \diff P\;\Phi(P),\qquad 1=\int \diff p\;\phi(p). 
\ee
\par
Let us briefly recall how different values of $\eta$ may arise in the context of the Langevin description of relativistic Brownian motions, as developed in \cite{DuHa05a}. 
Specializing to the rest frame of the heat bath, one can derive the following  Langevin equation for the stochastic motion of a relativistic Brownian particle \cite{DuHa05a}:  
\be\label{e:Langevin}
\f{\diff P}{\diff t}=-\nu P+ \left[\f{E_\eta(P)}{M}\right]^{1/2}\;\xi(t).
\ee
Here, $\nu$ is the friction parameter and $\xi$ ordinary Gaussian white noise with amplitude $\mcal{D}$, i.e., $\xi$ is characterized by 
\be
\lan \xi(t)\ran=0,\qquad
\lan \xi(t)\xi(t')\ran=2\mcal{D}\;\gd(t-t').
\ee
On the rhs. of Eq. \eqref{e:Langevin} the noise $\xi$ couples multiplicatively  to (a function of) the momentum coordinate $P$. Hence, different discretization rules may yield different stationary momentum distributions \cite{VK03}. It is convenient to parameterize discretization rules as follows
\be
E_\eta(P)=E[\eta\, P(t)+(1-\eta)\, P(t+dt)],\qquad \eta\in[0,1],
\ee
with $E(P)$ denoting the relativistic kinetic energy. In this notation, e.g., $\eta=0$ corresponds to the HK post-point discretization rule \cite{Han78,Han80,HaTo82,Kl94}, $\eta=1/2$ to the  
Fisk-Stratonovich mid-point discretization \cite{Fisk63,St64,St66} and
$\eta=1$ to Ito's pre-point discretization \cite{Ito44}. Introducing the temperature via the Einstein relation
\be
k_BT=\f{\mcal{D}}{m\nu},
\ee 
the candidate PDFs in Eqs.~\eqref{e:Rel_Maxwell} represent the stationary distributions associated with different values of $\eta$ \cite{DuHa05a}. In particular, only for $\eta=0$ (HK rule) the standard relativistic  J\"uttner-Maxwell distribution is recovered. 
\par
\begin{figure}[h]
\center
\epsfig{file=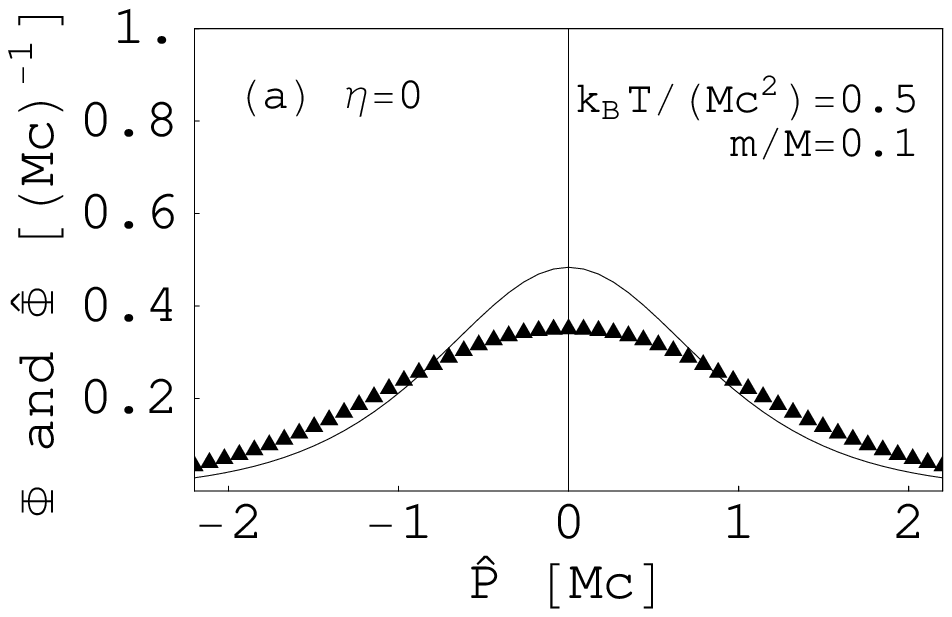,height=3.5cm, angle=-0}
\epsfig{file=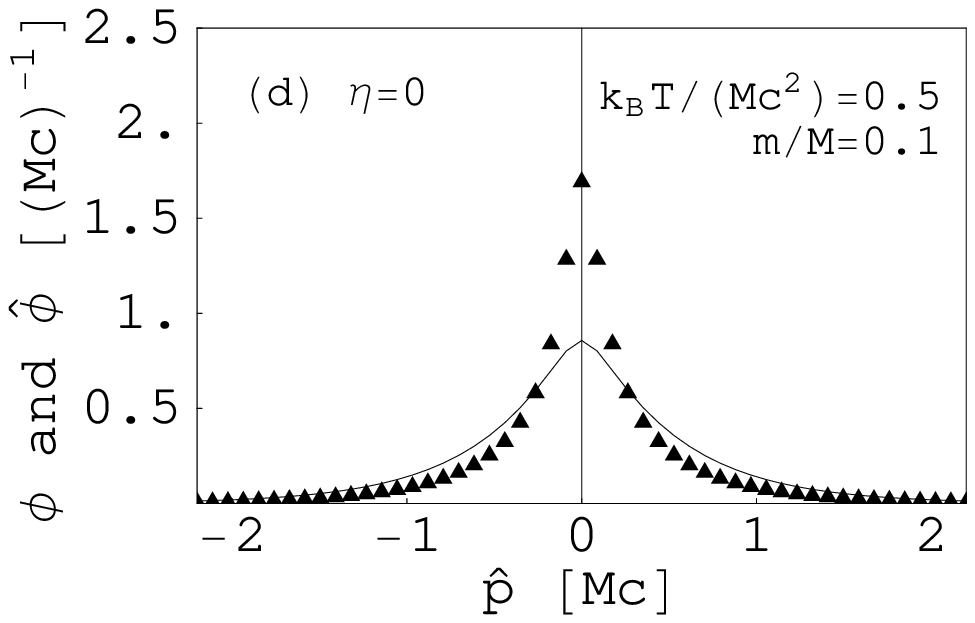,height=3.5cm, angle=-0}\\
\epsfig{file=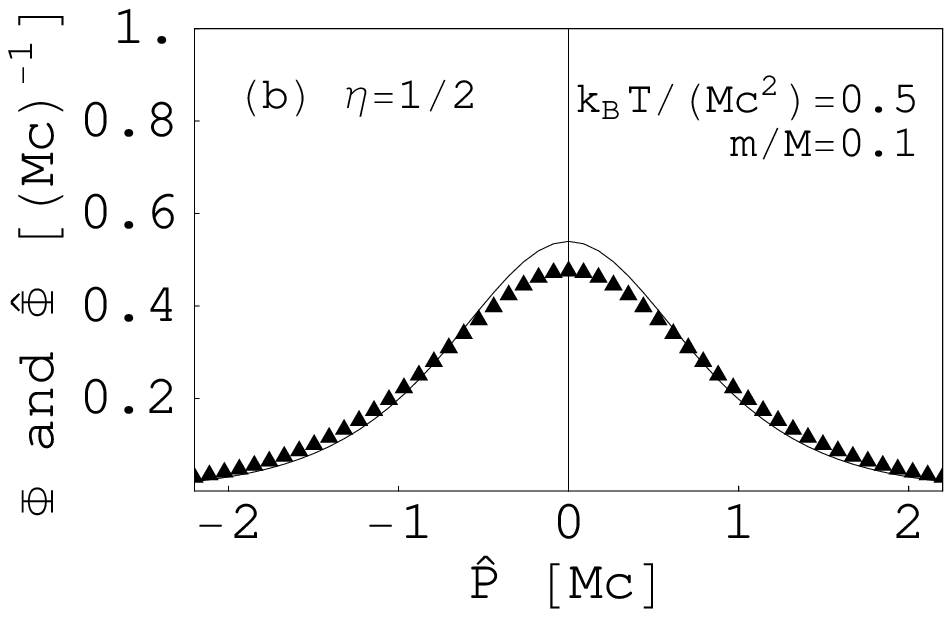,height=3.5cm, angle=-0}
\epsfig{file=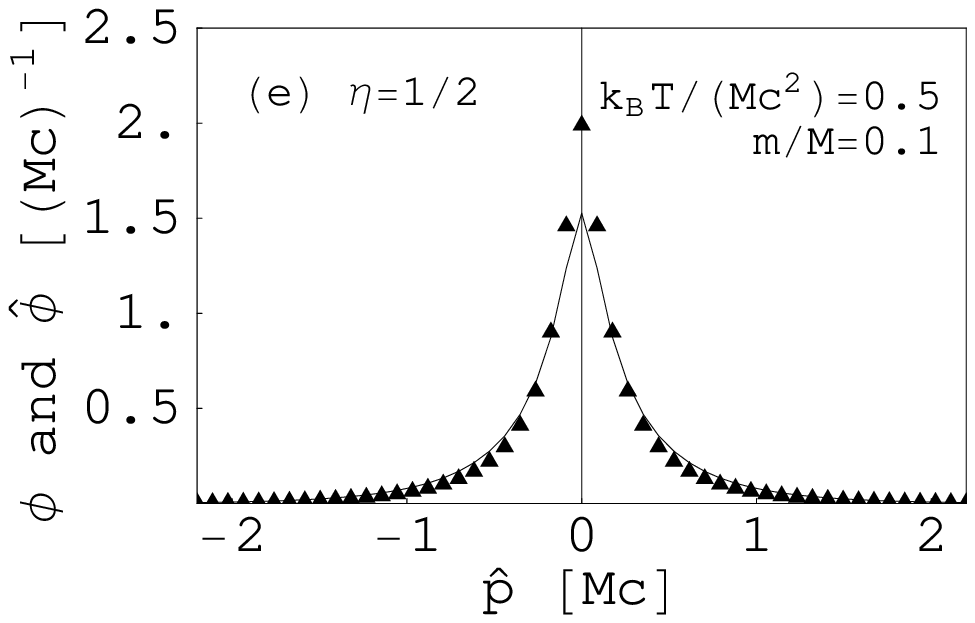,height=3.5cm, angle=-0}\\
\epsfig{file=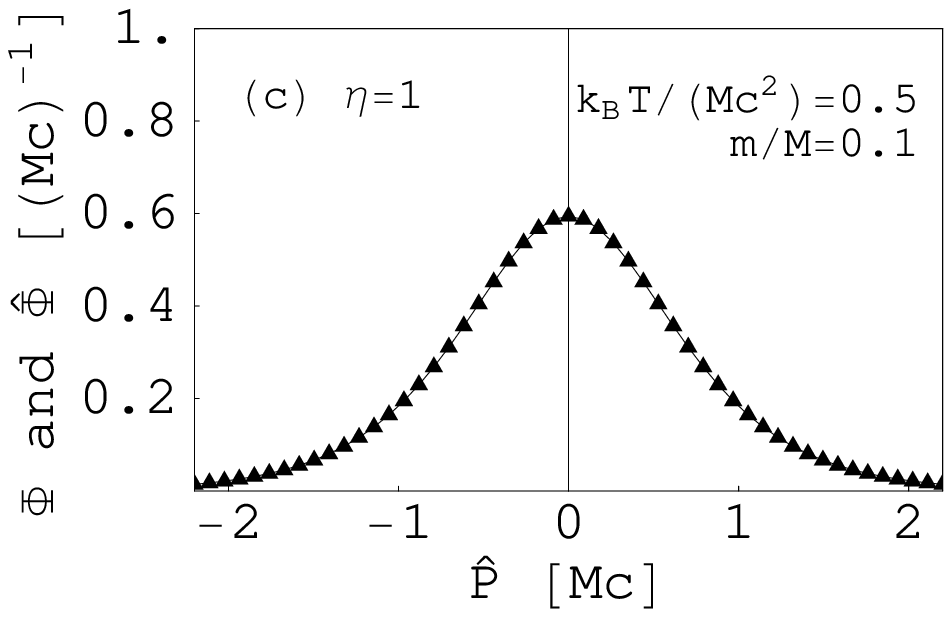,height=3.5cm, angle=-0}
\epsfig{file=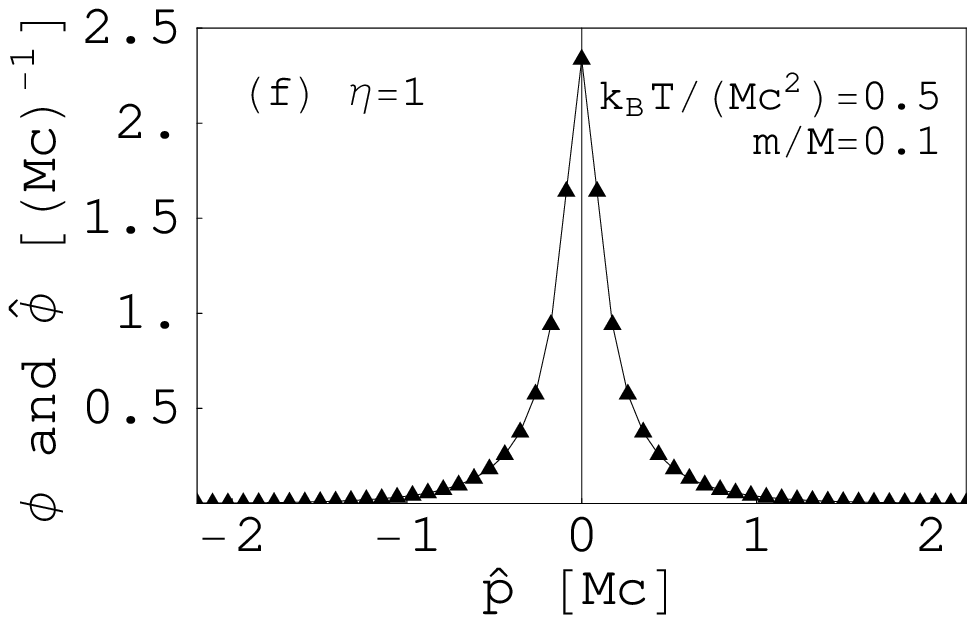,height=3.5cm, angle=-0}
\caption{\label{fig01}
Initial distributions (solid line) and numerical solutions (triangles) of Eqs.~\eqref{e:criteria_relativistic_used} for the momentum PDFs of the Brownian particle (a-c) and the heat bath particles (d-f). As one readily observes, see diagrams (c) and (f), only for $\eta=1$ the initial distributions are left invariant by the elastic collision process.}
\end{figure}

We are now in the position to discuss the numerical results. 
Figures \ref{fig01} (a-c) and \ref{fig01} (e-f) show functions
$\hat{\Phi}(\hat P)$ and $\hat{\phi}(\hat p)$, respectively, as
obtained for different values of $\eta$. In each diagram the solid
lines correspond to the initial momentum distributions from
Eqs. \eqref{e:Rel_Maxwell}. The triangles indicate the distributions
resulting after the collision, $\hat{\Phi}(\hat P)$ and
$\hat{\phi}(\hat p)$, obtained by numerically integrating
Eqs. \eqref{e:nonstationary} at 50 different values of $\hat P$ and
$\hat p$, respectively.  As one readily observes,  for $\eta=1$,
corresponding to  diagrams (c) and (f), the initial distributions
remain invariant in the course the elastic collision process. Hence,
according to these results, the modified  J\"uttner functions
\eqref{e:Rel_Maxwell} with $\eta=1$ are the relativistic analogs of
the the non-relativistic Maxwell distribution. For completeness, we
mention that we have tested the integral criteria over a wide range of
temperature and mass parameters and always found that only the
distributions with $\eta=1$ are invariant (all numerical integrations
were performed with the  function  \texttt{NIntegrate} of the computer
algebra program  {\sc Mathematica}~\cite{Ma03}). Furthermore, we note
that the normalization was conserved with high accuracy during the
numerical integration, i.e., the numerically found distributions
$(\hat\Phi,\hat{\phi})$ remained normalized to unity with high accuracy.    

\section{Summary and discussion}
\label{summary}

We have studied a simple microscopic model for 1D non-relativistic
and relativistic Brownian motions. It was our main objective to
identify the (invariant) stationary momentum distributions for both
Brownian and heat bath particles on the basis of the underlying
microscopic collision processes. To this end we have formulated two
integral criteria, which relate the initial momentum distributions to
the momentum distributions resulting after an elastic collision of a
Brownian particle with a heat bath particle (Sec.~\ref{s:integral}). 
The assumptions (postulates) underlying derivation of the integral
criteria can be summarized as follows:
\begin{itemize}
\item validity of the standard kinematic conservation laws, 
\item occurrence of the collision event, 
\item independence of the initial momenta.
\end{itemize}
It was then demonstrated that, under these assumptions, the integral criteria do correctly reproduce the Maxwellian distributions as the stationary solutions for the non-relativistic Brownian motions (Sec. \ref{s:Maxwell}). 
\par
Subsequently, the integral criteria were applied to the relativistic
case (Sec.~\ref{s:relativistic_criterion}). Here, we found that \emph{the standard J\"uttner distributions,
corresponding to $\eta=0$ in Eqs. \eqref{e:Rel_Maxwell}, are not
stationary with respect to the integral criteria}; i.e., given the
information that the collision has indeed occurred, the standard
J\"uttner distributions do not remain invariant in the course of an
elastic relativistic collision. Instead, \emph{the invariant
distributions are given by modified J\"uttner functions, corresponding
to $\eta=1$ in Eqs.~\eqref{e:Rel_Maxwell}}. This result is quite surprising,
since initially we had expected that the invariant solutions are
given by standard J\"uttner functions with $\eta=0$.  However, as
known from earlier work~\cite{DuHa05a}, modified J\"uttner
distributions with $\eta\ne 0$ may also appear as stationary solutions in
the 1D relativistic Langevin theory, with the value of $\eta$
depending on the discretization scheme that is used. In this context, we mention a recent paper by Lehmann~\cite{Le06}, who argues that J\"uttner's original approach \cite{Ju11} is non-covariant. Furthermore, we note that the invariant solution with $\eta=1$ can also be interpreted as a simple exponential (canonical) distribution) with respect to the Lorentz-invariant volume element of momentum space, $\diff^D p/p^0=\diff^D p/(p^2+m^2)^{1/2}$, where $D$ is the number of spatial dimensions \cite{SexlUrbantke}.    
\par
Due to the fact that our results are based on only three basic
assumptions, there is very little freedom for modifications such that
one could hope to recover the standard J\"uttner distributions as invariant
solutions. Hence, according to our opinion, this problem deserves
further consideration in the future. For example, as the next step, it
would be desirable to perform similar studies for simple 2D and 3D 
models~\footnote{Very recently, the authors were informed by L. O. Silva that Marti et al. have implemented a 3D relativistic Monte-Carlo model for charged particle collisions, and that their numerically obtained equilibrium solutions correspond to modified J\"uttner distributions with $\eta=1$ as well~(conference poster by M. Marti, R. A. Fonseca, L. O. Silva "A collision module for OSIRIS", P5.015, 32nd European Physical Society Conference on Plasma Physics, Rome, June 2006).}. Then the kinematics of a single collision process becomes more
complicated, because -- even for the simplest hard-sphere models --
momentum may be redistributed in different directions (depending on
the impact parameter). Hence, if one wishes to formulate analogous
integral criteria for identifying the stationary (invariant)  2D/3D
momentum distributions,  one will have to include additional equations
taking into account the cross-sections \footnote{We note, the
applicability of simple kinematic models as discussed here is, in
principle, limited to situations where high energies quantum
processes, as e.g.  creation and annihilation of particles, can be
neglected.}. However, if it should turn out that deviations from the
standard J\"uttner-Maxwell distribution persist in higher space
dimensions as well, then this might
be of relevance for calculating relativistic corrections in
high-energy physics \cite{2006PhRvC..73c4913V} and astrophysics (e.g., to the Sunyaev-Zeldovich
effect~\cite{SuZe72,ItKoNo98}). 

The authors are grateful to J. Chluba, A. Fingerle, and, in particular, to S. Hilbert for numerous helpful discussions.  


\par

\appendix
\section{Relativistic collisions}
\label{a:relativistic_collisions}
Our goal is to validate Eqs.~\eqref{e:p-tilde-rel}. To this end, consider two particles  having positions  $x$ and $X$ and relativistic momenta $p$ and $P$, as defined by Eq.~\eqref{e:relativistic_definitions}, with respect to (wrt.) the 1D inertial lab-frame $\Gs_0$. Assume that the particles collide at, say, $t=0$, which implies that:\\
(i) if $v(t)>V(t)$ at $t<0$, then $x(t)\le X(t)$ or, alternatively, \\
(ii) if $v(t)<V(t)$ at $t<0$, then $x(t)>X(t)$.\\ 
In  order calculate the momenta after the collision, $\hat{p}$ and
$\hat{P}$, as functions of the initial momenta, $p$ and $P$, it is
convenient to perform a Lorentz transformation to the center-of-mass frame. Before doing this, we briefly establish some notations. We define (1+1)-momenta wrt. $\Gs_0$ by
\be
\mfr{p}=(\epsilon,p),\qquad \mfr{P}=(E,P).
\ee
Assume that some inertial frame $\Gs'$ moves with velocity $v_0$ in $\Gs_0$, then the Lorentz transformation matrix $\Lambda_{v_0}$ can be parametrized as follows
\be
\Lambda_{v_0}=\f{1}{\sqrt{1-v_0^2}}\begin{pmatrix}
1&-v_0\\
-v_0&1
\end{pmatrix}
\ee
Its inverse is obtained by replacing $v_0$ with $-v_0$, i.e., 
\be
\Lambda^{-1}_{v_0}=\Lambda_{-v_0}=\f{1}{\sqrt{1-v_0^2}}\begin{pmatrix}
1&v_0\\
v_0&1
\end{pmatrix}.
\ee
For example, given $\mfr{P}=(E,p)$, the corresponding (1+1)-momentum vector wrt. $\Gs'$, denoted by $\mfr{P}'=(E',P')$, is obtained via matrix multiplication
\be
\mfr{P}'=\Lambda_{v_0}\,\mfr{P}.
\ee
 
\paragraph*{Center-of-mass frame. --}
In the following let us assume that $\Gs'$ is an inertial center-of-mass frame of the colliding particles; i.e., in $\Gs'$ we have, by definition,
\be
p'+P'=0.
\ee
This condition determines the Lorentz transformation parameter as
\be\label{a-e:v_0}
v_0=\f{p+P}{\epsilon+E}=\f{\hat{p}+\hat{P}}{\hat\epsilon+\hat E}=\hat{v}_0.
\ee
Furthermore, for elastic collisions, the energy and momentum balance become particularly simple in $\Gs'$:
\bse\label{a-e:balance}
\begin{align}
\hat{E}'&=E',\qquad\hat{\epsilon}'=\epsilon',\\
\hat{P}'&=-P'=p'=-\hat{p}',
\end{align} 
\ese
where, as before, hat-symbols refer to the momenta after the collision. 
It is convenient to express Eqs.~\eqref{a-e:balance} in matrix form, e.g., writing 
\bse\label{a-e:momentum_com}
\be
\hat{\mfr{P}}'=\gs'\,  \mfr{P}',\qquad
\hat{\mfr{p}}'=\gs'\,  \mfr{p}',
\ee
where the momentum transfer matrix $\gs'$ of the elastic collision is defined by
\be  
\gs'=\begin{pmatrix}
1&0\\
0&-1
\end{pmatrix}=\gs'^{-1}.
\ee
\ese

\paragraph*{Lab frame. --}
It is now straightforward to convert the above results \eqref{a-e:momentum_com} to the laboratory frame by applying the corresponding Lorentz transformations; e.g., for the Brownian particle we find
\bse
\be \label{a-e:momentum_BP}
\hat{\mfr{P}}
=\Lambda^{-1}_{v_0}\; \hat{\mfr{P}}'
=\Lambda^{-1}_{v_0}\; \gs'\;  \mfr{P}'
=\Lambda^{-1}_{v_0}\; \gs'\; \Lambda_{v_0}\; \mfr{P}
=\gs \; \mfr{P},
\ee
where
\be
\gs=\Lambda^{-1}_{v_0}\; \gs'\; \Lambda_{v_0}
=\f{1}{1-v_0^2}\begin{pmatrix}
1+v_0^2&-2v_0\\
2v_0&-(1+v_0^2)
\end{pmatrix}.
\ee
\ese
is the momentum transfer matrix wrt. $\Gs_0$, and $v_0$ is given by Eq.~\eqref{a-e:v_0}. Analogously, we find for the heat bath particle
\be \label{a-e:momentum_HP}
\hat{\mfr{p}}=\gs \, \mfr{p}.
\ee
An explicit evaluation of Eqs.~\eqref{a-e:momentum_BP} and \eqref{a-e:momentum_HP} yields
\bse
\be
\begin{pmatrix} \hat{E} \\ \hat P\end{pmatrix}
&=&\f{1}{1-v_0^2}
\begin{pmatrix}
(1+v_0^2)\, E-2 v_0\,P\\
2v_0\, E-(1+v_0^2)\,P
\end{pmatrix},\\
\begin{pmatrix} \hat{\epsilon} \\ \hat p\end{pmatrix}
&=&\f{1}{1-v_0^2}
\begin{pmatrix}
(1+v_0^2)\, \epsilon-2 v_0\,p\\
2v_0\,\epsilon-(1+v_0^2)\,p
\end{pmatrix},
\ee
\ese
which contains the desired result, cf. Eqs.~\eqref{e:p-tilde-rel}.
\par
Finally, we also calculate the inverse momentum transfer matrix
\be
\gs^{-1}=(\Lambda^{-1}_{v_0}\; \gs'\; \Lambda_{v_0})^{-1}
=\Lambda_{v_0}^{-1}\; {\gs'}^{-1}\; \Lambda_{v_0}
=\Lambda_{v_0}^{-1}\; \gs'\; \Lambda_{v_0}
=\gs,
\ee
allowing us to write
\bse
\be
\begin{pmatrix} {E} \\ P\end{pmatrix}
&=&\f{1}{1-\hat v_0^2}
\begin{pmatrix}
(1+\hat v_0^2)\, \hat E-2 \hat v_0\,\hat P\\
2\hat v_0\, \hat E-(1+\hat v_0^2)\,\hat P
\end{pmatrix},\\
\begin{pmatrix} {\epsilon} \\ p\end{pmatrix}
&=&\f{1}{1-\hat v_0^2}
\begin{pmatrix}
(1+\hat v_0^2)\,\hat \epsilon-2 \hat v_0\,\hat p\\
2\hat v_0\,\hat \epsilon-(1+\hat v_0^2)\,\hat p
\end{pmatrix}.
\ee
\ese

\end{document}